\documentclass[sigconf]{acmart}

\AtBeginDocument{%
  \providecommand\BibTeX{{%
    \normalfont B\kern-0.5em{\scshape i\kern-0.25em b}\kern-0.8em\TeX}}}

\copyrightyear{2020}
\acmYear{2020}
\setcopyright{acmlicensed}\acmConference[ESEM '20]{ESEM '20: ACM / IEEE International Symposium on Empirical Software Engineering and Measurement (ESEM)}{October 8--9, 2020}{Bari, Italy}
\acmBooktitle{ESEM '20: ACM / IEEE International Symposium on Empirical Software Engineering and Measurement (ESEM) (ESEM '20), October 8--9, 2020, Bari, Italy}
\acmPrice{15.00}
\acmDOI{10.1145/3382494.3410673}
\acmISBN{978-1-4503-7580-1/20/10}

\usepackage{bbding}
\usepackage{bm}
\usepackage{multirow}
\usepackage{tikz}
\usepackage{subcaption}
\usepackage{color}

\definecolor{light_green}{HTML}{DEE676}

\newcommand\encirclered[1]{%
  \tikz[baseline=(X.base)] 
    \node (X) [draw, shape=circle, inner sep=0, fill=red, text=white] {\strut #1};%
}
\newcommand\encirclelightgreen[1]{%
  \tikz[baseline=(X.base)] 
    \node (X) [draw, shape=circle, inner sep=0, fill=light_green, text=white] {\strut #1};%
}
\newcommand\encirclegreen[1]{%
  \tikz[baseline=(X.base)] 
    \node (X) [draw, shape=circle, inner sep=0, fill=green, text=white] {\strut #1};%
}

\begin{document}
\title{Long-Term Evaluation of Technical Debt in Open-Source Software}


\author{Arthur-Jozsef Molnar}
\email{arthur@cs.ubbcluj.ro}
\affiliation{%
  \institution{Babe\c{s}--Bolyai University}
  \streetaddress{M.Kogalniceanu 1}
  \city{Cluj-Napoca}
  \state{Romania}
  \postcode{400084}
}

\author{Simona Motogna}
\email{motogna@cs.ubbcluj.ro}
\affiliation{%
  \institution{Babe\c{s}--Bolyai University}
  \streetaddress{M.Kogalniceanu 1}
  \city{Cluj-Napoca}
  \state{Romania}
  \postcode{400084}
}


\begin{abstract}
\textbf{Background:} A consistent body of research and practice have identified that technical debt provides valuable and actionable insight into the design and implementation deficiencies of complex software systems. Existing software tools enable characterizing and measuring the amount of technical debt at selective granularity levels; by providing a computational model, they enable stakeholders to measure and ultimately control this phenomenon. \textbf{Aims:} In this paper we aim to study the evolution and characteristics of technical debt in open-source software. For this, we carry out a longitudinal study that covers the entire development history of several complex applications. The goal is to improve our understanding of how the amount and composition of technical debt changes in evolving software. We also study how new technical debt is introduced in software, as well as identify how developers handle its accumulation over the long term. \textbf{Method:} We carried out our evaluation using three complex, open-source Java applications. All 110 released versions, covering more than 10 years of development history for each application were analyzed using SonarQube. We studied how the amount, composition and history of technical debt changed during development, compared our results across the studied applications and present our most important findings. \textbf{Results:} For each application, we identified key versions during which large amounts of technical debt were added, removed or both. This had significantly more impact when compared to the lines of code or class count increases that generally occurred during development. However, within each version, we found high correlation between file lines of code and technical debt. We observed that the Pareto principle was satisfied for the studied applications, as 20\% of issue types generated around 80\% of total technical debt. Interestingly, there was a large degree of overlap between the issues that generated most of the debt across the studied applications. \textbf{Conclusions:} Early application versions showed greater fluctuation in the amount of existing technical debt. We found application size to be an unreliable predictor for the quantity of technical debt. Most debt was introduced in applications as part of milestone releases that expanded their feature set; likewise, we identified releases where extensive refactoring significantly reduced the level of debt. We also discovered that technical debt issues persist for a long time in source code, and their removal did not appear to be prioritized according to type or severity.
\end{abstract}



\maketitle

\section{Introduction}
Technical debt (TD) was introduced as a metaphor initially borrowed from the financial sector to express the debt that is accumulated when development of new features is prioritized over fixing known issues. The problem was first defined by Cunningham in 1992, who stated that a "\textit{little debt speeds development so long as it is paid back promptly with a rewrite}" \cite{29}. According to Fowler \cite{30}, technical debt represents deficiencies in the internal quality of software that make modifying and developing the system more difficult.

Similar to the financial one, technical debt has both a principal and an interest. The effort required to repay the original debt represents the principal, while the interest is the additional effort required to modify or extend the system caused by the presence of the principal. In this regard, as initially reported in \cite{29}, long-running debt incurs significant interest, which can overtake the principal and slow development to a crawl. Existing taxonomies categorize it into architectural, testing, source code as well as others manifestations of debt \cite{40}. Technical debt of all types can usually be addressed through refactoring. However, since the presence of debt is in many cases not directly noticed by end-users, time and budgetary pressures cause debt management to be postponed, as confirmed by several authors \cite{37,39,45}.   

\begin{table*}[h]
  \caption{Rule severity according to SonarQube documentation (\cite{1})}
  \label{tab:rule_severity}
  \begin{tabular}{cccp{10cm}}
    \toprule
    Severity & Impact & Likelihood & Description \\
    \midrule
    \multirow{2}{*}{Blocker} & \multirow{2}{*}{\CheckmarkBold} & \multirow{2}{*}{\CheckmarkBold} & High probability to impact the behavior of the application, such as a memory leak or unclosed JDBC connection. Code must be immediately fixed.\\
    \midrule
    \multirow{2}{*}{Critical} & \multirow{2}{*}{\CheckmarkBold} & \multirow{2}{*}{\XSolidBold} & Either a bug with a low probability to impact the behavior of the application or a security flaw such as an empty catch block or SQL injection.\\
    \midrule
    \multirow{2}{*}{Major} & \multirow{2}{*}{\XSolidBold} & \multirow{2}{*}{\CheckmarkBold} & Quality flaw which can highly impact developer productivity, such as an uncovered piece of code, duplicated blocks or unused parameters.\\
    \midrule
    \multirow{2}{*}{Minor} & \multirow{2}{*}{\XSolidBold} & \multirow{2}{*}{\XSolidBold} & Quality flaw which can slightly impact developer productivity, such as lines that are too long, or "switch" statements with fewer than 3 cases.\\
    \midrule
    \multirow{1}{*}{Info} & \multirow{1}{*}{\XSolidBold} & \multirow{1}{*}{\XSolidBold} & Findings that are not bugs or quality flaws, such as TODO's in code.\\
  \bottomrule
\end{tabular}
\end{table*}
While the research and practitioner communities have both addressed technical debt within the last decade, there still remain open issues. One of them regards the different perception that researchers, practitioners and management have on its importance and evaluation. This results in decisions relevant to debt management being informal and ad-hoc \cite{46}, with in many cases action taken only when development is severely impacted \cite{39}. This is compounded by the lack of a unified perspective regarding technical debt types, causes and impacts. While existing tools integrate the required components to measure technical debt, in many cases there are differences between reported results \cite{44,47}. Also, debt created early during the development cycle compounds interest and is more difficult to deal with, as evidenced by both earlier and recent research \cite{29,36}.

While literature covers both cross-sectional \cite{38,46,52} as well as longitudinal approaches \cite{4,39,51}, most of the existing studies do not cover the entire lifespan of the target applications. Furthermore, they do not include a fine-grained analysis of debt location, distribution and lifespan. Our study is aimed at filling this gap, while providing all the required steps to replicate our results or further extend them. We employ the SonarQube platform that is widely used in both research and industry and cover the full development history of three complex, popular open-source applications. The main contributions of our approach consist in (i) providing a long-term view on the evolution of technical debt in complex, open-source applications; (ii) studying technical debt composition and lifespan; (iii) studying the relation between application maturity and the presence of technical debt and (iv) identifying trends that are common between, or particular to the studied applications.

\section{Preliminaries}
\subsection{Technical debt}
From the software development point of view, technical debt represents an indicator that delivery time was prioritized over internal code quality \cite{30}. From a project management perspective, it is an indicator of otherwise hidden technical quality issues, which might realize their impact only later during development or maintenance. Since its importance was understood by researchers and practitioners alike, several tools were created to assist developers and other stakeholders to measure and control it. These include the SonarQube platform, Squore\footnote{\url{https://www.vector.com/int/en/products/products-a-z/software/squore/}}, Kiuwan\footnote{\url{https://www.kiuwan.com/}} or Ndepend\footnote{\url{www.ndepend.com}}. Our study is built around the SonarQube platform described in the following section.

Each tool is built around a quality model that measures technical debt. SonarQube, Squore and NDepend use the \textit{SQALE} \cite{31} model, while Kiuwan uses the \textit{Checking Quality Model}, which takes into consideration an evaluation of maintainability, reliability, portability, efficiency, and security. Other well-known models include \textit{QMOOD}, \textit{Quamoco} and \textit{Columbus}, which were examined in existing literature \cite{44,47}. Usually, software tools employ static code analysis in order to collect data about several metrics and code coverage that are verified against a set of rules. Evaluations are made according to the quality models to determine whether rules are satisfied, and determine the impact and cost of rule violations.

\subsection{SonarQube model}
\label{sec:sonarqube_mode}
SonarQube is used to monitor source code quality and security. Its community edition is free and open source. It includes the required plugins to provide analysis for 15 languages including Java, Python, C\#, XML and HTML. Supporting new languages is possible by creating plug-ins, an example of which is the open-source plugin for C++\footnote{\url{https://github.com/SonarOpenCommunity/sonar-cxx}}. Plugins add rule definitions for source code checked during analysis. When a rule is broken, a source code issue which derives its characteristics from the rule is created.

Issue characteristics important in our evaluation are the type, severity and effort required to fix. Type is one of \emph{code smell} (maintainability domain), \emph{bug} (reliability domain) or \emph{vulnerability} (security domain)\footnote{\url{https://docs.sonarqube.org/latest/user-guide/rules/}}. 

Severity is assigned according to the risk estimation procedure illustrated in Table \ref{tab:rule_severity}. Issues are also assigned one or more tags inherited from the rules that raised them. Rules also provide an estimation function for calculating the time required to fix generated issues. These usually provide either a constant time per issue, or a function with linear offset. For instance, rule \emph{java:S3776} states that \textit{"Cognitive Complexity of methods should not be too high"}. It generates code smells of critical severity with a linear time to fix, consisting of a constant 5 minute time per issue, plus 1 minute for each additional complexity point over an established threshold. 

As part of the analysis, SonarQube calculates the target system's technical debt, which is the estimated time required to fix all detected issues. SonarQube also computes the technical debt ratio $TDR = \frac{TD}{DevTime}$. $DevTime$ is the estimated time required to develop the system, with 1 line of code (LOC) estimated to cost 30 minutes of development. The $TDR$ is then classified on a SQALE rating between A (best, $TDR < 5\%$) and E (worst, $TDR \geq 50\%$). This provides a high-level, evidence-backed view of the system's internal quality.

Our evaluation employed SonarQube 8.2, which includes an updated version of the Java code analyzer, enhancing semantic rule resolution. Code is compiled using the Eclipse Java compiler and analysis is based on checking compliance with over 550 source code rules.

\section{Related Work}
A survey of publications between the first mention of technical debt (1992, \cite{29}) and 2013 \cite{40} showed the field attracted contributions equally from academia (43\%), industry (40\%), or in collaboration (17\%). An approach to estimate technical debt \cite{41} based on repair and maintenance effort was included in the SIG quality model \cite{48}. The estimation was based on data collected from 44 systems and intended to assist managers in analysing returns on investment. Another industrial study \cite{37} focused on identifying the sources and impact of debt using a small focus group. Interviews conducted with experienced professionals concluded that technical debt was in many cases not properly used in decision making. A case study \cite{44} involving 10 releases of 10 open source systems used correlation and regression analysis to confirm the impact of technical debt on maintainability, but for other software quality factors, the influence remained dependant on estimation technique. 

Researchers also carried out comparative evaluations for well-known quality models. Authors of \cite{47} noted the existence of inconsistencies between \textit{SQALE} and \textit{Quamoco}, as the two models remained in agreement only for software with \textit{A}-grade quality. When applied on systems having lower maintainability, reliability or security ratings, evaluations diverged. Authors of \cite{53} compared the SIG quality model \cite{48} against the maintainability index \cite{57} and SQALE for technical debt identification in a case study using 17 large-scale, open-source Python libraries. Results showed differences between the debt calculated according to the models, with more similarity reported between the maintainability index and the SIG model. Longitudinal evaluation showed most of the major software releases increased the level of technical debt; generally, later software versions accumulated more technical debt across all studied methods. Authors of \cite{44} carried out an empirical evaluation showing discrepancies between technical debt estimation and software quality according to \textit{QMOOD}. 

In \cite{52}, authors carried out an empirical analysis on 91 Apache Java systems from several domains including big data, database, network server and web frameworks. A positive relation was identified between the number of commits, system size and technical debt within the same software domain. Authors noted higher debt density in smaller-sized systems, as well as differences between systems that belonged to different software domains. In \cite{39}, authors evaluated the technical debt specific to system architecture within 5 large companies. They identified its most prevalent causes, including uncertainty in early version use cases and continued focus on feature delivery over architecture. Complete, partial and no refactoring were examined as possible strategies to handle architectural technical debt. Complete refactoring was not practical within the confines of the study due to frequently changing requirements and pressure to deliver expected features. Conversely, ignoring technical debt lead to frequent crises, where development of new features slowed down or was halted until debt was addressed. Based on study data and interviews, authors concluded that partial refactoring provided the best balance between the involved trade-offs. The relation between reported defects and reliability issues reported by SonarQube were examined in a longitudinal case study that covered the evolution of 21 open-source software systems in the span of 39,500 commits \cite{55}. Several machine learning methods were used to determine the most fault-prone SonarQube violations using 4,500 defects that were mapped to commits. The study showed only 26 of the 200+ SonarQube Java rules were related with the introduction of software defects. Authors concluded that SonarQube reliability issues did not directly generate software defects, and that further work was required to translate SonarQube issue characterization to bug reports for production software.

A longitudinal case study focused on identifying and paying architectural debt was detailed in \cite{51}. Focusing on a healthcare communications product, its commit history was employed to identify the existence of structural problems that motivated further architectural analysis. Issues such as groups of strongly connected components, violations of modular design and improper use of inheritance were used as markers of architectural debt. The software was evaluated for 6 months before and 6 months after a major architectural refactoring. Results showed that successfully addressing architectural debt improved the time required to resolve issues by 72\% and nearly doubled the number of defects resolved per month.

Existing literature also covers the long-term effects of technical debt accumulation. Authors of \cite{54} reported the results from a longitudinal study covering 10 months, during which 43 developers supplied more then 470 reports. Results showed developers wasted 23\% of their time handing technical debt, as well as being forced to knowingly add new debt due to time pressure and existing debt. Most time was wasted on additional testing, source code analysis and required refactoring. Ghanbari et al. \cite{56} explored the psychological effect of technical debt on developers through interview and survey. They showed developers were familiar with technical debt and were not comfortable with taking on more of it, as the expectation of additional work in the future lowered their motivation. Respondents showed more interest in repaying debt, which was seen as "\textit{an interesting activity, which motivates them as well as enables them to constantly increase the quality of software artefacts and feel confident}" \cite{56}. Developers reported cases where fixing debt was dangerous due to the possibility of introducing faults in an otherwise working system. While developers were not criticized for introducing additional debt, they felt that companies also did not reward timely debt management sufficiently.

As most of the studies are empirical, more evidence needs to be reported in order to improve our understanding of technical debt issues. The results presented in this study are differentiated from related work through the coverage of the complete development history, and the analysis of the composition and lifespan of technical debt for several target applications.

\begin{table*}
    \caption{Details for earliest and latest application versions in our study}
    \label{tab:target_applications}
    \centering
        \begin{tabular}{ccccrrrc}
        \toprule
        \multirow{2}{*}{Application} & \multirow{2}{*}{Version} & \multirow{2}{*}{Release Date} & \multirow{2}{*}{LOC} & \multicolumn{3}{c}{Open Issues (SQALE rating)} & Technical Debt \\
        &  &  & & Bugs & Vulnerabilities & Code Smells & (work days) \\
        \midrule
        \multirow{2}{*}{FreeMind} & 0.0.3 & July 9, 2000 & 2,770 & 1 \multirow{2}{*}{\encirclered{\small{E}}}  & 10 \encirclelightgreen{\small{B}} & 185 \multirow{2}{*}{\encirclegreen{\small{A}}} & 5 \\
        & 1.1.0Beta2 & Feb 5, 2016 & 43,269 & 76 \hphantom{\encirclered{\small{E}}} & 208 \encirclered{\small{E}} & 3,386 \hphantom{\encirclegreen{\small{A}}} & 86 \\ 
        \midrule        
        \multirow{2}{*}{jEdit} & 2.3pre2 & Jan 29, 2000 & 22,311 & 48 \multirow{2}{*}{\encirclered{\small{E}}} & 59 \encirclelightgreen{\small{B}} & 1,177 \multirow{2}{*}{\encirclegreen{\small{A}}} & 43 \\
        & 5.5.0 & April 9, 2018 & 96,195 & 145 \hphantom{\encirclered{\small{E}}} & 253 \encirclered{\small{E}} & 9,249 \hphantom{\encirclegreen{\small{A}}} & 211 \\
        \midrule
        \multirow{2}{*}{TuxGuitar} & 0.1pre & June 18, 2006 & 8,960 & 34 \multirow{2}{*}{\encirclered{\small{E}}} & 88 \multirow{2}{*}{\encirclered{\small{E}}} & 868 \multirow{2}{*}{\encirclegreen{\small{A}}} & 11 \\
        & 1.5.3 & Dec 10, 2019 & 105,113 & 163 \hphantom{\encirclered{\small{E}}} & 286 \hphantom{\encirclered{\small{E}}} & 3,758 \hphantom{\encirclegreen{\small{A}}} & 77 \\
        \bottomrule
        \end{tabular}
\end{table*}

\section{Case Study}
\subsection{Research Objective}
\label{sec:research_objective}
We structured our evaluation according to the best practices detailed by Runeson and Höst \cite{17}. The main objective of our paper stated using the goal-question-metric \cite{18} is to \textit{"investigate the presence, characteristics and long-term evolution of source code technical debt in open-source software"}. We operationalize our objective in the form of the following research questions:

\bm{$RQ_{1}$}: \textit{When and where is technical debt introduced in source code?}
Time and budget constraints have been identified as some of the root causes for technical debt \cite{37}. Empirical studies have also shown architectural changes and refactoring efforts to influence technical debt more than software size \cite{3,39}. The objective is to improve our understanding of how and where debt is introduced in application source code. When considering software defects, we find the Pareto principle applicable in many cases. A recent empirical study \cite{38} demonstrated that a small number of files were accountable for most of the defects in large systems. We investigate whether a similar relation can be identified for technical debt.

\bm{$RQ_{2}$}: \textit{What is the composition of source code technical debt?}
We aim to improve our understanding of how maintainability, reliability and security issues are represented in overall technical debt. Furthermore, we are interested in providing a finer-grained characterization by breaking technical debt down by severity, which can assist practitioners in prioritizing the allocation of resources to address critical issues. Finally, we drill down to SonarQube rule level in order to examine how each rule contributes to overall technical debt, verify whether the Pareto principle applies and if affirmative, attempt a formal characterization.

\bm{$RQ_{3}$}: \textit{How does technical debt evolve over the long term?}
We identify software versions that are key for debt management, as previous research showed that versions exist where many issues are introduced or resolved \cite{3,4}. In addition, research has confirmed that early versions are prone to software changes that impact maintainability and security. We examine whether such trends can be identified for technical debt in our target applications, and we compare how debt type, severity and composition evolve over the long-term. In addition, we aim to report on issue lifespan and study whether developers prioritize fixing issues based on type, severity or associated tags. 

\subsection{Target Applications}
\label{sec:target_applications}
Previous research \cite{52} has shown that target software domain influences technical debt characteristics. To ensure data triangulation \cite{17} and enable cross-application analysis for $RQ_{2}$ and $RQ_{3}$, we restrict our case study to GUI-driven applications on the Java platform. We acknowledge that empirical research showed open-source development to be the subject of hiatuses \cite{49}. In addition, large-scale cross-sectional analyses illustrated that in many cases, repository source code was incomplete, or produced compilation or run-time errors \cite{2}. We minimized these risks by focusing our search on mature applications having a fully documented development history and a well established user base. This allowed us to study the evolution of technical debt both in early versions, which existing research has shown to be more unstable \cite{2,3}, as well as including mature versions, where the assumption is that application architecture is well established. The main exclusion criteria was the existence of dependencies on external components such as databases, devices or an Internet connection, as these were expected to make curating target applications more difficult without benefiting our study.

The selected target applications are the FreeMind\footnote{\url{http://freemind.sourceforge.net/wiki/index.php/Main\_Page}} mind mapping software, the jEdit \footnote{\url{http://jedit.org}} text-file editor and the TuxGuitar\footnote{\url{http://www.tuxguitar.com.ar}} multi-track tablature manager. Table \ref{tab:target_applications} provides details about application sizes and quality issues for the earliest and latest version of each application in our study. Throughout the paper we employ the developer assigned version numbers, as the staggered development rhythm of many open source projects, together with several development hiatuses lead us to conclude that using moments or duration in time was inappropriate. For instance, FreeMind 0.9.0Beta17 was released less than 2 months after version 0.8.1 and is a major update at both functionality and source code level. In contrast, the duration separating versions 0.8.0 and 0.8.1, which are very similar is around 2\textonehalf{ } years. We acknowledge that a staggered release plan caused by development taking place on different branches is a possible explanation, but accounting for this remains beyond our scope.

The \textbf{FreeMind} mind-mapper integrates plugin support and was the target of previous software engineering research \cite{5}. The application has an important user base, with over 577k downloads over the last year\footnote{All download data points taken on May $19^{th} 2020$.}. As shown in Table \ref{tab:target_applications}, its earliest version is the smallest application in our study. Consistently, we found early FreeMind versions less complex when compared to those of jEdit or TuxGuitar. Figure \ref{fig:application_size_td} shows a steady increase in application size and a large spike in versions 0.8.0 and 0.8.1, after which LOC increases very slowly. 

The \textbf{jEdit} text editor has a consistent user base, exceeding 90k yearly downloads. It caters to programmers, has an integrated plugin manager and a large collection of open-source plugins. As shown in Figure \ref{fig:application_size_td}, it has a large number of releases, which consistently increased application size. We found all released versions, including the earliest ones to be feature-rich and stable. jEdit was also used in testing \cite{5,6}, object-oriented metric \cite{7} and maintainability \cite{3} research.

The \textbf{TuxGuitar} multi-track tablature manager is the third application in our study. Its evolution appears similar to jEdit, with each release increasing the size of its code base, as shown in Figure \ref{fig:application_size_td}. It is perhaps the most active of the studied applications, with over 250k downloads recorded during the last year and the most recent version released in December, 2019.

\subsection{Methodology}
\label{sec:methodology}
\subsubsection{Data collection} We downloaded the source code for all application releases. We handled the case of several releases happening in the span of a few days by including only the most recent of them in our study, to keep the number of versions manageable. This resulted in 38 FreeMind, 45 jEdit and 27 TuxGuitar releases included in our study. The source code for each of the 110 versions was imported into an IDE and manually examined. Several versions packaged library and application code together, such as the \textit{com.microstar.xml} parser or the \textit{BeanShell} interpreter included in several jEdit releases. This code was compiled separately and added to the classpath. We also ensured that FreeMind and jEdit releases did not include plugin code. For TuxGuitar, we found that key functionalities were implemented using developer-created plugins and decided to include this code in our evaluation. We ran each included release and manually checked that all functionalities were working as expected.

\begin{figure*}
    \captionsetup[subfigure]{labelformat=empty,justification=centering}
    \centering
    \begin{subfigure}[b]{\textwidth}
        \includegraphics[width=\textwidth]{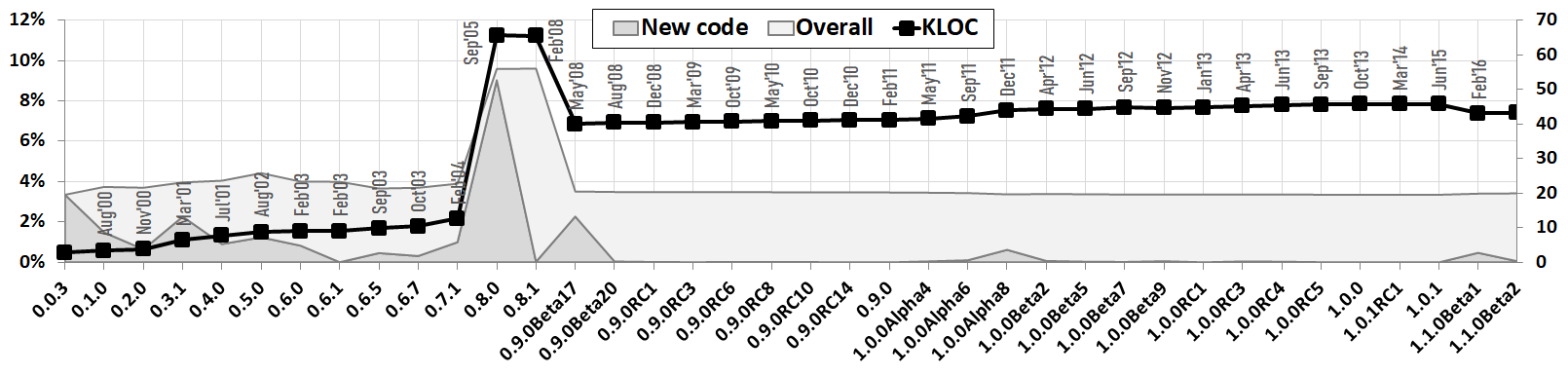}
    \end{subfigure}
    \begin{subfigure}[b]{\textwidth}
        \includegraphics[width=\textwidth]{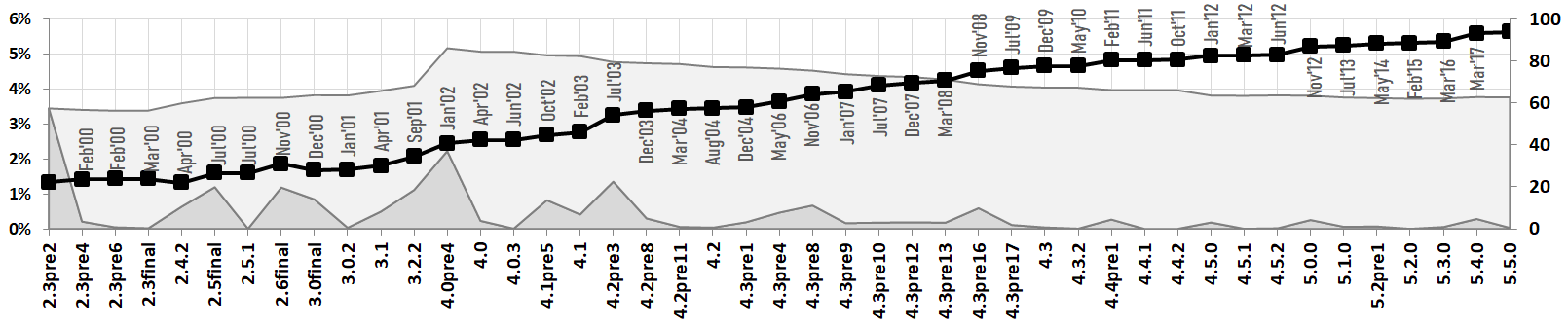}
    \end{subfigure}
    \par\medskip
    \begin{subfigure}[b]{\textwidth}
        \includegraphics[width=\textwidth]{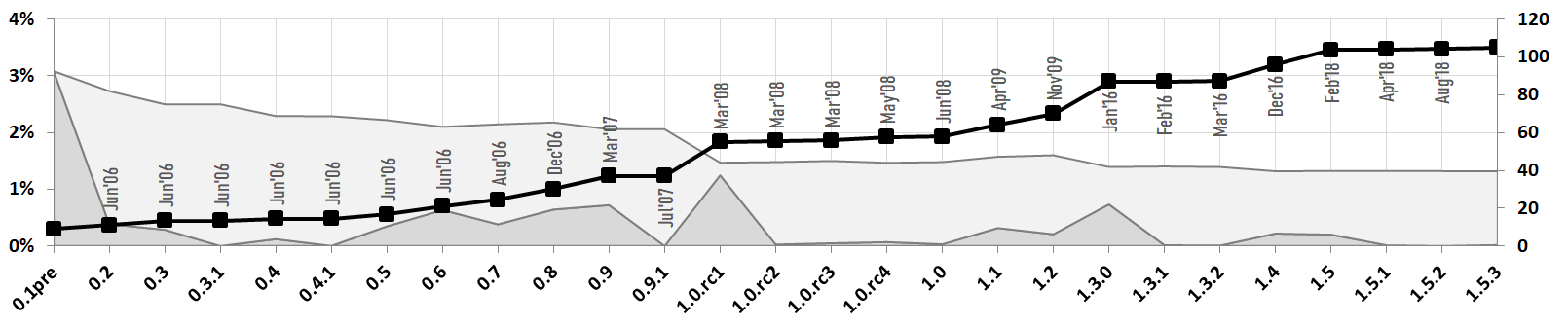}
    \end{subfigure}
    \caption{Information about studied versions of FreeMind (top), jEdit (middle) and TuxGuitar (bottom). Size information according to SonarQube's \textit{ncloc} metric uses the scale on the right; each version's release date is also shown. The hashed area shows overall $TDR$ (Section \ref{sec:sonarqube_mode}) for each version using light gray, and debt newly added in that version using darker gray}
    \label{fig:application_size_td}
\end{figure*}

\subsubsection{SonarQube analysis}\label{sec:sq_analysis} All released versions were imported into a SonarQube instance that was configured to support long-running projects\footnote{\url{https://docs.sonarqube.org/latest/analysis/analysis-parameters/}}. This entailed configuring the server to keep project history permanently, importing projects in ascending version order and setting the new code window to cover the time span between the previous and current versions. These server-side changes allowed SonarQube to track the history of each created issue and to automatically close issues no longer detected in later versions. This is shown in Figure \ref{fig:application_size_td}, which illustrates  the technical debt introduced in each version. SonarQube analysis resulted in 54,617 unique issues, estimated to generate 1,316 work-days worth of technical debt. 37,093 of these issues, worth 916 work-days were fixed within the analyzed versions. 

Data analysis was carried out using purpose-written Python scripts that use the SonarQube API to extract, aggregate and process project data. To facilitate replicating or extending our evaluation, we open-sourced the issue database together with our intermediate and final results \cite{50}.

\subsection{Results}
\label{sec:results}
We present our most important results, seen through the lens of the research questions defined in Section \ref{sec:research_objective}. 

\subsubsection{\bm{$RQ_{1}$}: When and where is technical debt introduced in source code?}
We used the SonarQube data recorded for each application version. In addition, for each version we calculated newly introduced debt, which allowed us to determine key versions, during which an important part of debt was added or removed. We used the $TDR$ normalization of technical debt according to software size. We used statement lines of code\footnote{The \textit{ncloc} SonarQube metric} as the proxy for application size; we carried out a Spearman test that uncovered very high correlation ($>0.98$) between this metric and the number of application classes and methods for all three target applications.

Figure \ref{fig:application_size_td} illustrates application size in statement lines of code ($KLOC$) and the $TDR$ for each version. All technical debt is new in the first version of each application. As $TDR$ is normalized to application size, deviation from a horizontal line represents supplemental debt that was introduced or eliminated. For each application, we identified key versions, where large swaths of debt were introduced, eliminated, or both. We discuss these in more detail when answering $RQ_{3}$. One observation common to the target applications is that latter versions, such as FreeMind versions after 0.9.0Beta17 proved more stable debt-wise. They no longer introduced, nor eliminated as much debt, and all of them received an \textit{A}-rating according to the SQALE model.

As existing research has identified correlation between file size and defects \cite{38}, we carried out a Spearman rank correlation between technical debt and source file LOC for each studied version. Results showed very strong correlation for FreeMind ($mean=0.86$, standard deviation $\sigma=0.07$) and jEdit ($mean=0.89$, $\sigma=0.04$) and strong for TuxGuitar ($mean=0.74$, $\sigma=0.09$). In all studied versions, top 20\% files according to LOC carried over half the technical debt. We found only some application versions satisfied Pareto's law, so we stopped short of attempting to determine a power law to describe the relation between technical debt and file size, such as in \cite{38}. Likewise, the bottom 20\% smallest files contained very little debt, in many cases less than 5\% of total. We concluded that similar to software defects, most technical debt resided in large files; however this did not provide additional information regarding severity, composition or lifespan.

\begin{table}
  \caption{Distribution of the effort required to fix issues according to type. Reported as mean \% values across all application versions, together with standard deviation. For example, the mean time required to fix all FreeMind vulnerabilities is 4.6\% of the effort required to fix all issues.}
    \begin{tabular}{rrrr}
    \multirow{2}{*}{Type} & FreeMind & jEdit & TuxGuitar\\
    & \% ($\sigma$) & \% ($\sigma$) & \% ($\sigma$)\\
    \midrule
    Bug & 2.8 (1.1) & 1.4 (0.3) & 5.8 (1.9)\\
    Vulnerability & 4.6 (1.0) & 2.8 (0.6) & 9.5 (2.2)\\
    Code smell & 92.6 (1.0) & 95.8 (0.5) & 84.7 (3.5)\\
    \bottomrule
    \end{tabular}
  \label{tab:issue_type}%
\end{table}

\subsubsection{\bm{$RQ_{2}$}: What is the composition of source code technical debt?}
Tables \ref{tab:issue_type} and \ref{tab:issue_severity} illustrate the mean distribution of issue type and severity according to total remediation effort. We found most debt was reported for the maintainability domain (code smells), while fixing reliability issues (bugs) would take the least effort. This remained consistent across target applications, as well as across their history, as shown by the low values of standard deviation. A similar hierarchy occurred for issue severity, where blocker and critical issues were in the minority. 

\begin{table}
  \caption{Distribution of the effort required to fix issues according to severity. Reported as mean \% values across all application versions, together with standard deviation.}
    \begin{tabular}{rrrr}
    \multirow{2}{*}{Severity} & FreeMind & jEdit & TuxGuitar\\
    & \% ($\sigma$) & \% ($\sigma$) & \% ($\sigma$)\\
    \midrule
    Blocker & 1.2 (0.3) & 0.6 (0.1) & 2.4 (1.4)\\
    Critical & 21.4 (3.8) & 11.5 (1.4) & 10.6 (2.8)\\
    Major & 57.5 (6.1) & 74.5 (2.8) & 61.8 (1.9)\\
    Minor & 19.1 (4.3) & 11.8 (2.7) & 25.0 (3.3)\\
    \bottomrule
    \end{tabular}
  \label{tab:issue_severity}%
\end{table}

\begin{table*}
  \caption{Minimal set of rules generating >50\% technical debt effort in each of the three applications. Rules generating critical issues in bold. Rule ID, tags and description from SonarQube 8.2}
  \label{tab:rules_50}
  \begin{tabular}{p{3.55cm}cccl}
    Rule ID (SonarQube tags) & FreeMind & jEdit & TuxGuitar & Description \\
    \toprule
    S110 (design) & 16.8\% & 34.8\% & 1.1\% & Inheritance tree of classes should not be too deep\\
    dupl. (pitfall) & 10.8\% & 1.5\% & 22.1\% & Source files should not have any duplicated blocks\\
    S125 (unused) & 4.5\% & 23.1\% & 1.9\% & Sections of code should not be commented out\\
    \textbf{S3776 (brain-overload)} & 2.8\% & 4.2\% & 9.1\% & \textbf{Cognitive Complexity of methods should not be too high}\\
    \textbf{S1948 (serialization, cwe)} & 10.5\% & 3.9\% & 0.4\% & \textbf{Fields in a "Serializable" class should be transient or serializable}\\
    S1149 (performance) & 7.1\% & 4.1\% & 0.6\% & Synchronized classes [...] should not be used\\
    S1874 (cert, cwe, obsolete) & 4.1\% & 3.9\% & 1.7\% & "@Deprecated" code should not be used\\
    S1181 (bad-practice, cert, cwe, error-handling) & \multirow{2}{*}{0.04\%} & \multirow{2}{*}{0.8\%} & \multirow{2}{*}{7.8\%} & \multirow{2}{*}{Throwable and Error should not be caught}\\
    S1854 (cert, cwe, unused) & 1.4\% & 0.5\% & 5.3\% & Unused assignments should be removed\\
\midrule
\textbf{TOTAL} & \textbf{58.04}\% & \textbf{76.8}\% & \textbf{50}\% \\
\bottomrule
\end{tabular}
\end{table*}

\begin{table}
  \caption{Minimal set of tags that account for >80\% technical debt issues in each of the three applications.}
  \label{tab:tags_80}
  \begin{tabular}{rcccl}
    SonarQube tag & FreeMind & jEdit & TuxGuitar \\
    \midrule
    cwe & 17.40\% & 13.47\% & 19.33\% \\
    cert & 18.87\% & 15.09\% & 14.21\% \\
    unused & 13.36\% & 18.67\% & 13.67\% \\
    bad-practice & 6.71\% & 8.32\% & 8.86\% \\
    convention & 12.99\% & 6.03\% & 1.82\% \\
    clumsy & 4.30\% & 3.92\% & 5.08\% \\
    suspicious & 2.56\% & 3.74\% & 4.61\% \\
    obsolete & 1.79\% & 8.15\% & 0.75\% \\
    pitfall & 2.04\% & 2.99\% & 5.10\% \\
    brain-overload & 2.52\% & 3.14\% & 3.04\% \\
    performance & 2.19\% & 4.69\% & 1.59\% \\
    confusing & 4.17\% & 2.21\% & 1.26\% \\
    error-handling & 1.13\% & 1.60\% & 4.84\% \\
    \midrule
\textbf{TOTAL} & \textbf{90.01}\% & \textbf{92.01}\% & \textbf{84.17}\% \\
\bottomrule
\end{tabular}
\end{table}

We examined the most prevalent issue types according to their  tag\footnote{\url{https://docs.sonarqube.org/latest/user-guide/built-in-rule-tags/}} and rule information, as shown in Tables \ref{tab:rules_50} and \ref{tab:tags_80}. The first observation is that around 80\% of reported issues were grouped under one third of the tags. These are broken down for each application in Table \ref{tab:tags_80}. We noticed significant overlap, with tags describing security-related issues such as \textit{cwe} (Common Weakness Enumeration)\footnote{\url{http://cwe.mitre.org/}} and \textit{cert} (SEI CERT Coding Standards)\footnote{\url{https://wiki.sei.cmu.edu/confluence/display/java}} prominent in all three applications. We also examined the source of the discrepancies between the applications. For instance, the \textit{convention} tag was associated with many FreeMind issues due to variables not following the Java naming scheme in older application code. Likewise, the prevalence of issues marked as \textit{obsolete} in jEdit was caused by the existence or use of code previously marked as deprecated.

We also examined the data at rule level to better break down the composition of debt. For jEdit, we found that 20\% of rules covered >80\% of debt, satisfying the Pareto principle. This did not happen for most versions of FreeMind or TuxGuitar. We illustrate using Table \ref{tab:rules_50} the smallest common set of 9 rules that generate >50\% of total technical debt in all three applications. We observed rules that include issues of architectural debt (\textit{S110}, \textit{S1874}), complexity (\textit{S3776}) and reliability (\textit{S1181}).

\subsubsection{\bm{$RQ_{3}$}: How does technical debt evolve over the long term?}
Figure \ref{fig:application_size_td} illustrates how overall debt evolved over time. It also helps identify key versions where important changes took place. In the case of FreeMind, 0.8.0 was a pivotal version that added many new functionalities, updated the UI and increased application size from 12.5 KLOC to 65.5 KLOC. These changes were accompanied by the introduction of over 370 days of added technical debt, which lowered its \textit{SQALE} maintainability rating from \textit{A} to \textit{B}. Most of this debt was fixed in version 0.9.0Beta17, which also lowered application size without sacrificing functionalities. Releases after 0.9.0Beta17 were much more stable, both in terms of application size as well as regarding technical debt, and retained the \textit{A} maintainability rating. Conversely, we found more fluctuation in versions earlier to 0.8.0, where smaller application sizes compounded the weight of introduced issues in the calculation of the $TDR$.

jEdit version 4.0pre4 was also interesting to examine, as new functionalities were associated with increased technical debt. Versions released after 4.0 illustrated a gradual descent of the debt ratio, with little new debt being added. This was coupled with a steady and expected increase in application size, showing that newly added code was of high quality.

The evolution of TuxGuitar coupled a gradual decrease in technical debt with a large increase of the code base. We single out version 1.0rc1 for several aspects. First, it introduced new features such as the song collection browser and additional plugins. Also, many code smells were resolved in this version, further lowering an already excellent $TDR$. However, newly added source code also contributed 42 days worth of debt, some of which was fixed in latter versions.

Figure \ref{fig:application_size_td} also uncovers many versions during which technical debt levels remained stable, so we extended our examination to study the lifespan of detected issues. As detailed in Section \ref{sec:sq_analysis}, SonarQube automatically closes those issues that can no longer be detected in newer application versions. This is either the result of refactoring, or a side-effect of other source code level changes. As these changes occurred between major releases, a commit-level analysis would be required to correctly characterize them. 

\begin{figure*}
    \captionsetup[subfigure]{labelformat=empty,justification=centering}
    \centering
    \begin{subfigure}[b]{0.33\textwidth}
        \includegraphics[width=\textwidth]{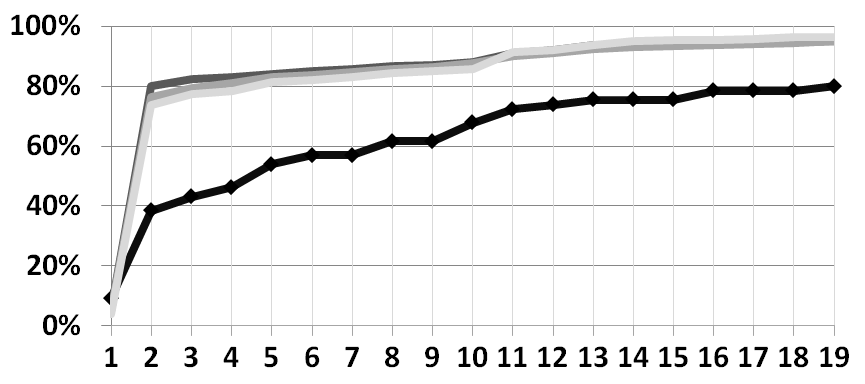}
    \end{subfigure}
    \begin{subfigure}[b]{0.33\textwidth}
        \includegraphics[width=\textwidth]{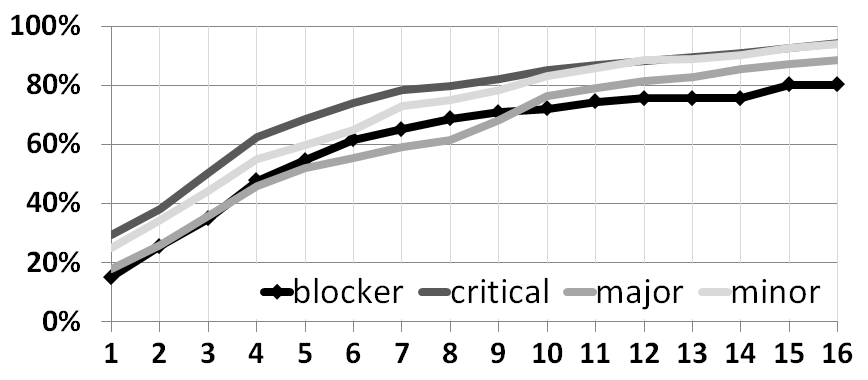}
    \end{subfigure}
    \begin{subfigure}[b]{0.33\textwidth}
        \includegraphics[width=\textwidth]{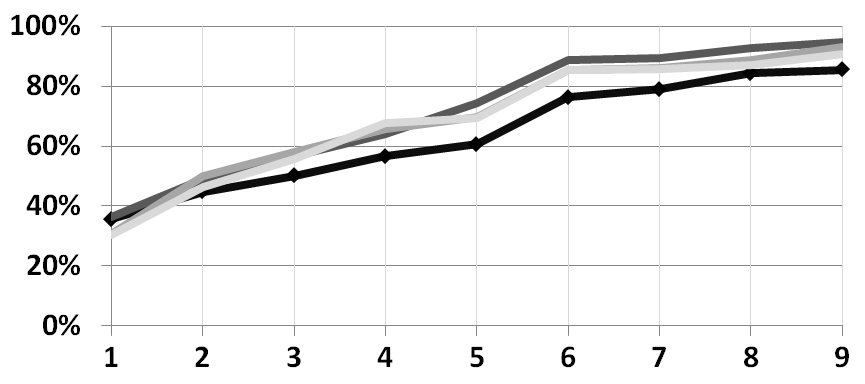}
    \end{subfigure}
    \caption{Number of releases required to fix 80\% of technical debt issues according to their severity in FreeMind (left), jEdit (center) and TuxGuitar (right).}\label{fig:versions_severity}
\end{figure*}

\begin{figure*}
    \captionsetup[subfigure]{labelformat=empty,justification=centering}
    \centering
    \begin{subfigure}[b]{0.33\textwidth}
        \includegraphics[width=\textwidth]{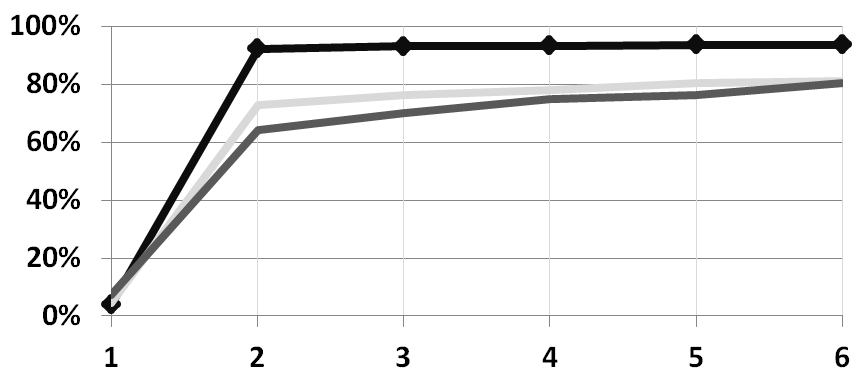}
    \end{subfigure}
    \begin{subfigure}[b]{0.33\textwidth}
        \includegraphics[width=\textwidth]{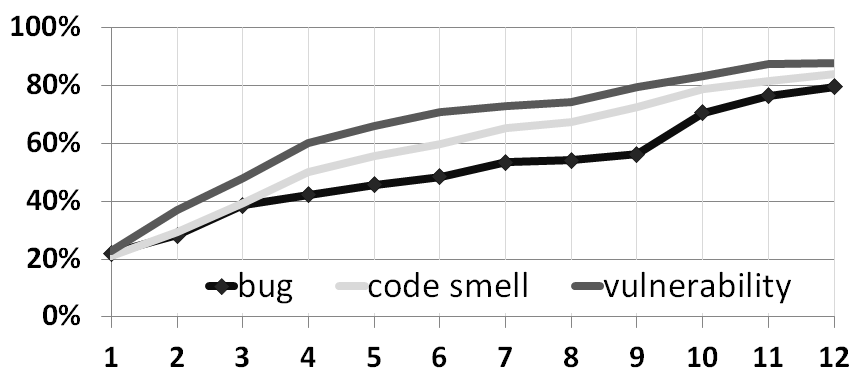}
    \end{subfigure}
    \begin{subfigure}[b]{0.33\textwidth}
        \includegraphics[width=\textwidth]{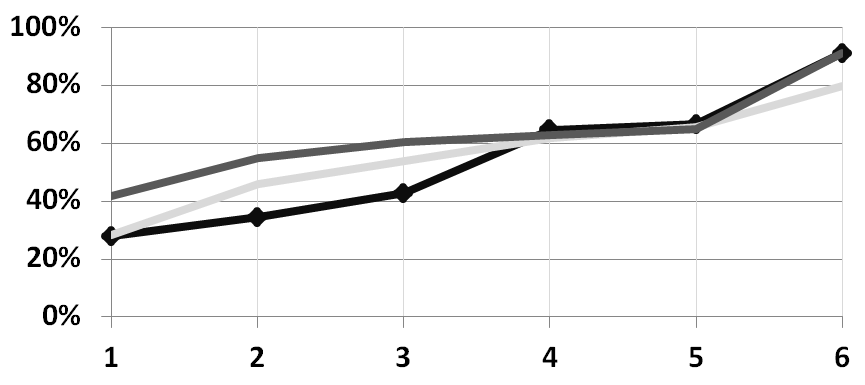}
    \end{subfigure}
    \caption{Number of releases required to fix 80\% of technical debt issues according to their type in FreeMind (left), jEdit (center) and TuxGuitar (right).}\label{fig:versions_type}
\end{figure*}

We studied issue lifespan in terms of version releases, and not according to time. This was due to existing research that identified that work on many open-source projects was riddled with key person departures and development hiatuses \cite{49}. Figure \ref{fig:application_size_td} shows our target systems were no exception. Figures \ref{fig:versions_severity} and \ref{fig:versions_type} show the proportion of issues fixed per application according to severity and type, respectively. Most FreeMind issues were fixed within 2 releases, which is skewed by the large spike corresponding to technical debt added in version 0.8.0 and removed two releases later. For jEdit and TuxGuitar, we noted a more gentle slope, with 80\% of all detected issues being fixed only after 16 and 9 releases, respectively. Our data showed that resolving issues did not seem to be prioritized according to issue type nor severity. Furthermore, like in our answer for $RQ_{2}$, we examined the data according to associated tags. Again, there was no correlation between issue lifespan and associated tags.

\subsection{Threats to Validity}
\label{sec:threats}
We organized our evaluation according to the best practices defined in \cite{17}. The major steps we carried out were to define the research objective, select target applications, collect, process and analyze the data. Our work includes a publicly-available replication package \cite{50} that facilitates verifying our results or extending the study. 

\subsubsection{Internal threats}
We automated the data analysis process and included safeguards and tests in our source code. We externalized the analysis to spreadsheet software, with all analysis results manually checked for correctness. We used SonarQube's web interface to check and confirm each step of our process. Data in raw, intermediate and final forms are publicly available \cite{50}. The most important remaining threat regards the applicability of the SonarQube model. Research \cite{47,53} has shown that differences exist in the evaluation of software quality between existing models. Furthermore, in \cite{55} authors showed that many of the reliability issues reported by SonarQube did not actually lead to observable faults.

\subsubsection{External threats}
We consider the selection of target applications the most important external threat. Selecting applications similar in type and development platform improves study consistency and enables data triangulation; however, it also produces results that cannot be easily generalized to other system types. This is further evidenced in \cite{52}, where authors noted the existence of differences in the characterization of technical debt between Apache suite applications that belonged to different software domains. Since differences in software tooling, target applications and study methodology make direct comparison across software domains impossible, we must caution against extrapolating the reported results without careful consideration. 

Additionally, we expect the observer bias that is introduced by widespread use of code quality platforms such as SonarSource\footnote{\url{https://www.sonarsource.com/}} to change the profile of current and future target applications. Coupled with the observations from Lenarduzzi et al. \cite{55}, this might result in developers placing more emphasis on achieving good static analysis ratings in the detriment of fixing defects.

\subsubsection{Construct threats}
We limited our evaluation to issues detected by the most recent version of SonarQube. Furthermore, we considered a semantic examination of generated issues, or an interpretation of their importance beyond the scope of the present paper. While SonarQube does report issue severity and a list of associated tags, not all issues translate into functionality bugs or an actionable attack surface. Likewise, there might be important issues that remain undetected when limiting the evaluation to static analysis. In our analysis, we employed technical debt effort represented by the duration required to fix detected issues. However, there exists the possibility that the platform over or under-reports these durations, skewing the results of our analysis.

\section{Conclusions}
We employed the SonarQube platform for code quality and security measurement to evaluate technical debt in three complex open-source applications. We carried out a longitudinal study in order to improve our understanding of the characteristics and lifespan of technical debt over the long term. We use the change in technical debt data to identify key software versions. We confirm that early application versions are less stable when compared to their mature counterparts, as suggested by previous research \cite{3,7,29}. 

We identified versions where significant debt was both introduced as well as removed. One such example is FreeMind version 0.8.0, where the $TDR$ increased by 5.5\%. Figure \ref{fig:application_size_td} reveals most of this debt was new, and that part of the debt initially identified in version 0.7.1 was resolved. A similar situation occurred in TuxGuitar 1.0rc1, which we highlighted to showcase the effects of software refactoring on existing technical debt. Again, Figure \ref{fig:application_size_td} revealed the overall decrease in debt to be the result of opposite actions. On one hand, refactoring reduced the debt carried from previous versions; on the other, new debt was also introduced. This illustrates that a detailed examination is required to identify technical debt variance, and that limiting analyses to the $TDR$ can be misleading. Next, we carried out a fine-grained analysis on debt composition. We showed that a small number of rules accounted for most of the effort required to reduce technical debt; this leads us to believe that many underlying issues could be fixed by improvements in the planning and design phases, as proposed by existing research \cite{29,37,39}.

Our current goal is to extend the analysis to applications from other software domains, as well as to include commercial applications with a traceable development history. We wish to corroborate source code analysis with developer feedback and improve our understanding of the rationale behind our findings. Furthermore, we aim to extend our study in order to include technical lag \cite{58} and study its relation with technical debt at both a fundamental level as well as through the lens of open-source software.

\bibliographystyle{ACM-Reference-Format}
\bibliography{references}


\begin{thebibliography}{33}


\ifx \showCODEN    \undefined \def \showCODEN     #1{\unskip}     \fi
\ifx \showDOI      \undefined \def \showDOI       #1{#1}\fi
\ifx \showISBNx    \undefined \def \showISBNx     #1{\unskip}     \fi
\ifx \showISBNxiii \undefined \def \showISBNxiii  #1{\unskip}     \fi
\ifx \showISSN     \undefined \def \showISSN      #1{\unskip}     \fi
\ifx \showLCCN     \undefined \def \showLCCN      #1{\unskip}     \fi
\ifx \shownote     \undefined \def \shownote      #1{#1}          \fi
\ifx \showarticletitle \undefined \def \showarticletitle #1{#1}   \fi
\ifx \showURL      \undefined \def \showURL       {\relax}        \fi
\providecommand\bibfield[2]{#2}
\providecommand\bibinfo[2]{#2}
\providecommand\natexlab[1]{#1}
\providecommand\showeprint[2][]{arXiv:#2}

\bibitem[\protect\citeauthoryear{Alfayez, Chen, Behnamghader, Srisopha, and
  Boehm}{Alfayez et~al\mbox{.}}{2018}]%
        {52}
\bibfield{author}{\bibinfo{person}{Reem Alfayez}, \bibinfo{person}{Celia Chen},
  \bibinfo{person}{Pooyan Behnamghader}, \bibinfo{person}{Kamonphop Srisopha},
  {and} \bibinfo{person}{Barry Boehm}.} \bibinfo{year}{2018}\natexlab{}.
\newblock \showarticletitle{An Empirical Study of Technical Debt in Open-Source
  Software Systems}. In \bibinfo{booktitle}{\emph{Disciplinary Convergence in
  Systems Engineering Research}}, \bibfield{editor}{\bibinfo{person}{Azad~M.
  Madni}, \bibinfo{person}{Barry Boehm}, \bibinfo{person}{Roger~G. Ghanem},
  \bibinfo{person}{Daniel Erwin}, {and} \bibinfo{person}{Marilee~J. Wheaton}}
  (Eds.). \bibinfo{publisher}{Springer International Publishing},
  \bibinfo{address}{Cham}, \bibinfo{pages}{113--125}.
\newblock
\showISBNx{978-3-319-62217-0}


\bibitem[\protect\citeauthoryear{Arlt, Banerjee, Bertolini, Memon, and
  Schaf}{Arlt et~al\mbox{.}}{2012}]%
        {5}
\bibfield{author}{\bibinfo{person}{Stephan Arlt}, \bibinfo{person}{Ishan
  Banerjee}, \bibinfo{person}{Cristiano Bertolini}, \bibinfo{person}{Atif~M.
  Memon}, {and} \bibinfo{person}{Martin Schaf}.}
  \bibinfo{year}{2012}\natexlab{}.
\newblock \showarticletitle{Grey-box GUI Testing: Efficient Generation of Event
  Sequences}.
\newblock \bibinfo{journal}{\emph{CoRR}}  \bibinfo{volume}{abs/1205.4928}
  (\bibinfo{year}{2012}).
\newblock


\bibitem[\protect\citeauthoryear{{Avelino}, {Constantinou}, {Valente}, and
  {Serebrenik}}{{Avelino} et~al\mbox{.}}{2019}]%
        {49}
\bibfield{author}{\bibinfo{person}{G. {Avelino}}, \bibinfo{person}{E.
  {Constantinou}}, \bibinfo{person}{M.~T. {Valente}}, {and} \bibinfo{person}{A.
  {Serebrenik}}.} \bibinfo{year}{2019}\natexlab{}.
\newblock \showarticletitle{On the abandonment and survival of open source
  projects: An empirical investigation}. In \bibinfo{booktitle}{\emph{2019
  ACM/IEEE International Symposium on Empirical Software Engineering and
  Measurement (ESEM)}}. \bibinfo{pages}{1--12}.
\newblock


\bibitem[\protect\citeauthoryear{Barkmann, Lincke, and Löwe}{Barkmann
  et~al\mbox{.}}{2009}]%
        {2}
\bibfield{author}{\bibinfo{person}{H. Barkmann}, \bibinfo{person}{R. Lincke},
  {and} \bibinfo{person}{W. Löwe}.} \bibinfo{year}{2009}\natexlab{}.
\newblock \showarticletitle{Quantitative Evaluation of Software Quality Metrics
  in Open-Source Projects}. In \bibinfo{booktitle}{\emph{2009 International
  Conference on Advanced Information Networking and Applications Workshops}}.
  \bibinfo{pages}{1067--1072}.
\newblock
\urldef\tempurl%
\url{https://doi.org/10.1109/WAINA.2009.190}
\showDOI{\tempurl}


\bibitem[\protect\citeauthoryear{Besker, Martini, and Bosch}{Besker
  et~al\mbox{.}}{2018}]%
        {54}
\bibfield{author}{\bibinfo{person}{Terese Besker}, \bibinfo{person}{Antonio
  Martini}, {and} \bibinfo{person}{Jan Bosch}.}
  \bibinfo{year}{2018}\natexlab{}.
\newblock \showarticletitle{Technical Debt Cripples Software Developer
  Productivity: A Longitudinal Study on Developers’ Daily Software
  Development Work}. In \bibinfo{booktitle}{\emph{Proceedings of the 2018
  International Conference on Technical Debt}} (Gothenburg, Sweden)
  \emph{(\bibinfo{series}{TechDebt ’18})}. \bibinfo{publisher}{Association
  for Computing Machinery}, \bibinfo{address}{New York, NY, USA},
  \bibinfo{pages}{105–114}.
\newblock
\showISBNx{9781450357135}
\urldef\tempurl%
\url{https://doi.org/10.1145/3194164.3194178}
\showURL{%
\tempurl}


\bibitem[\protect\citeauthoryear{Caldiera and Rombach}{Caldiera and
  Rombach}{1994}]%
        {18}
\bibfield{author}{\bibinfo{person}{Victor R Basili1~Gianluigi Caldiera} {and}
  \bibinfo{person}{H~Dieter Rombach}.} \bibinfo{year}{1994}\natexlab{}.
\newblock \showarticletitle{{The Goal Question Metric approach}}.
\newblock \bibinfo{journal}{\emph{Encyclopedia of software engineering}}
  (\bibinfo{year}{1994}), \bibinfo{pages}{528--532}.
\newblock


\bibitem[\protect\citeauthoryear{Ghanbari, Besker, Martini, and Bosch}{Ghanbari
  et~al\mbox{.}}{2017}]%
        {56}
\bibfield{author}{\bibinfo{person}{Hadi Ghanbari}, \bibinfo{person}{Terese
  Besker}, \bibinfo{person}{Antonio Martini}, {and} \bibinfo{person}{Jan
  Bosch}.} \bibinfo{year}{2017}\natexlab{}.
\newblock \showarticletitle{Looking for Peace of Mind? Manage Your (Technical)
  Debt: An Exploratory Field Study}. In \bibinfo{booktitle}{\emph{Proceedings
  of the 11th ACM/IEEE International Symposium on Empirical Software
  Engineering and Measurement}} (Markham, Ontario, Canada)
  \emph{(\bibinfo{series}{ESEM ’17})}. \bibinfo{publisher}{IEEE Press},
  \bibinfo{pages}{384–393}.
\newblock
\showISBNx{9781509040391}
\urldef\tempurl%
\url{https://doi.org/10.1109/ESEM.2017.53}
\showURL{%
\tempurl}


\bibitem[\protect\citeauthoryear{Gonzalez-Barahona, Sherwood, Robles, and
  Izquierdo}{Gonzalez-Barahona et~al\mbox{.}}{2017}]%
        {58}
\bibfield{author}{\bibinfo{person}{Jesus~M. Gonzalez-Barahona},
  \bibinfo{person}{Paul Sherwood}, \bibinfo{person}{Gregorio Robles}, {and}
  \bibinfo{person}{Daniel Izquierdo}.} \bibinfo{year}{2017}\natexlab{}.
\newblock \showarticletitle{Technical Lag in Software Compilations: Measuring
  How Outdated a Software Deployment Is}. In \bibinfo{booktitle}{\emph{Open
  Source Systems: Towards Robust Practices}},
  \bibfield{editor}{\bibinfo{person}{Federico Balaguer},
  \bibinfo{person}{Roberto Di~Cosmo}, \bibinfo{person}{Alejandra Garrido},
  \bibinfo{person}{Fabio Kon}, \bibinfo{person}{Gregorio Robles}, {and}
  \bibinfo{person}{Stefano Zacchiroli}} (Eds.). \bibinfo{publisher}{Springer
  International Publishing}, \bibinfo{address}{Cham},
  \bibinfo{pages}{182--192}.
\newblock


\bibitem[\protect\citeauthoryear{{Griffith}, {Reimanis}, {Izurieta}, {Codabux},
  {Deo}, and {Williams}}{{Griffith} et~al\mbox{.}}{2014}]%
        {44}
\bibfield{author}{\bibinfo{person}{I. {Griffith}}, \bibinfo{person}{D.
  {Reimanis}}, \bibinfo{person}{C. {Izurieta}}, \bibinfo{person}{Z. {Codabux}},
  \bibinfo{person}{A. {Deo}}, {and} \bibinfo{person}{B. {Williams}}.}
  \bibinfo{year}{2014}\natexlab{}.
\newblock \showarticletitle{The Correspondence Between Software Quality Models
  and Technical Debt Estimation Approaches}. In \bibinfo{booktitle}{\emph{2014
  Sixth International Workshop on Managing Technical Debt}}.
  \bibinfo{pages}{19--26}.
\newblock


\bibitem[\protect\citeauthoryear{Henrik}{Henrik}{2013}]%
        {36}
\bibfield{author}{\bibinfo{person}{Kniberg Henrik}.}
  \bibinfo{year}{2013}\natexlab{}.
\newblock \bibinfo{title}{Good and Bad Technical Debt (and how TDD helps)}.
\newblock
\newblock
\urldef\tempurl%
\url{https://blog.crisp.se/2013/10/11/henrikkniberg/good-and-bad-technical-debt}
\showURL{%
\tempurl}


\bibitem[\protect\citeauthoryear{{Izurieta}, {Griffith}, and
  {Huvaere}}{{Izurieta} et~al\mbox{.}}{2017}]%
        {47}
\bibfield{author}{\bibinfo{person}{C. {Izurieta}}, \bibinfo{person}{I.
  {Griffith}}, {and} \bibinfo{person}{C. {Huvaere}}.}
  \bibinfo{year}{2017}\natexlab{}.
\newblock \showarticletitle{An Industry Perspective to Comparing the SQALE and
  Quamoco Software Quality Models}. In \bibinfo{booktitle}{\emph{2017 ACM/IEEE
  International Symposium on Empirical Software Engineering and Measurement
  (ESEM)}}. \bibinfo{pages}{287--296}.
\newblock


\bibitem[\protect\citeauthoryear{Klinger, Tarr, Wagstrom, and Williams}{Klinger
  et~al\mbox{.}}{2011}]%
        {46}
\bibfield{author}{\bibinfo{person}{Tim Klinger}, \bibinfo{person}{Peri Tarr},
  \bibinfo{person}{Patrick Wagstrom}, {and} \bibinfo{person}{Clay Williams}.}
  \bibinfo{year}{2011}\natexlab{}.
\newblock \showarticletitle{An Enterprise Perspective on Technical Debt}. In
  \bibinfo{booktitle}{\emph{Proceedings of the 2nd Workshop on Managing
  Technical Debt}} (Waikiki, Honolulu, HI, USA) \emph{(\bibinfo{series}{MTD
  ’11})}. \bibinfo{publisher}{Association for Computing Machinery},
  \bibinfo{address}{New York, NY, USA}, \bibinfo{pages}{35–38}.
\newblock
\showISBNx{9781450305860}
\urldef\tempurl%
\url{https://doi.org/10.1145/1985362.1985371}
\showURL{%
\tempurl}


\bibitem[\protect\citeauthoryear{{Lenarduzzi}, {Lomio}, {Huttunen}, and
  {Taibi}}{{Lenarduzzi} et~al\mbox{.}}{2020}]%
        {55}
\bibfield{author}{\bibinfo{person}{V. {Lenarduzzi}}, \bibinfo{person}{F.
  {Lomio}}, \bibinfo{person}{H. {Huttunen}}, {and} \bibinfo{person}{D.
  {Taibi}}.} \bibinfo{year}{2020}\natexlab{}.
\newblock \showarticletitle{Are SonarQube Rules Inducing Bugs?}. In
  \bibinfo{booktitle}{\emph{2020 IEEE 27th International Conference on Software
  Analysis, Evolution and Reengineering (SANER)}}. \bibinfo{pages}{501--511}.
\newblock
\urldef\tempurl%
\url{https://arxiv.org/abs/1907.00376}
\showURL{%
\tempurl}


\bibitem[\protect\citeauthoryear{Lenarduzzi, Orava, Saarimaki, Systa, and
  Taibi}{Lenarduzzi et~al\mbox{.}}{2019}]%
        {37}
\bibfield{author}{\bibinfo{person}{V. Lenarduzzi}, \bibinfo{person}{T. Orava},
  \bibinfo{person}{N. Saarimaki}, \bibinfo{person}{K. Systa}, {and}
  \bibinfo{person}{D. Taibi}.} \bibinfo{year}{2019}\natexlab{}.
\newblock \showarticletitle{An Empirical Study on Technical Debt in a Finnish
  SME}. In \bibinfo{booktitle}{\emph{2019 ACM/IEEE International Symposium on
  Empirical Software Engineering and Measurement (ESEM)}}.
  \bibinfo{publisher}{IEEE Computer Society}, \bibinfo{address}{Los Alamitos,
  CA, USA}, \bibinfo{pages}{1--6}.
\newblock
\urldef\tempurl%
\url{https://doi.ieeecomputersociety.org/10.1109/ESEM.2019.8870169}
\showURL{%
\tempurl}


\bibitem[\protect\citeauthoryear{Letouzey}{Letouzey}{2012}]%
        {31}
\bibfield{author}{\bibinfo{person}{Jean-Louis Letouzey}.}
  \bibinfo{year}{2012}\natexlab{}.
\newblock \showarticletitle{The SQALE Method for Evaluating Technical Debt}. In
  \bibinfo{booktitle}{\emph{Proceedings of the Third International Workshop on
  Managing Technical Debt}} \emph{(\bibinfo{series}{MTD '12})}.
  \bibinfo{publisher}{IEEE Press}, \bibinfo{pages}{31--36}.
\newblock
\showISBNx{978-1-4673-1749-8}
\urldef\tempurl%
\url{http://dl.acm.org/citation.cfm?id=2666036.2666042}
\showURL{%
\tempurl}


\bibitem[\protect\citeauthoryear{Li, Avgeriou, and Liang}{Li
  et~al\mbox{.}}{2014}]%
        {40}
\bibfield{author}{\bibinfo{person}{Zengyang Li}, \bibinfo{person}{Paris
  Avgeriou}, {and} \bibinfo{person}{Peng Liang}.}
  \bibinfo{year}{2014}\natexlab{}.
\newblock \showarticletitle{A Systematic Mapping Study on Technical Debt and
  Its Management}.
\newblock \bibinfo{journal}{\emph{Journal of Systems and Software}}
  (\bibinfo{date}{12} \bibinfo{year}{2014}), \bibinfo{pages}{193--220}.
\newblock
\urldef\tempurl%
\url{https://doi.org/10.1016/j.jss.2014.12.027}
\showDOI{\tempurl}


\bibitem[\protect\citeauthoryear{Martin}{Martin}{2019a}]%
        {45}
\bibfield{author}{\bibinfo{person}{Fowler Martin}.}
  \bibinfo{year}{2019}\natexlab{a}.
\newblock \bibinfo{title}{Is High Quality Software Worth the Cost?}
\newblock
\newblock
\urldef\tempurl%
\url{https://martinfowler.com/articles/is-quality-worth-cost.html}
\showURL{%
\tempurl}


\bibitem[\protect\citeauthoryear{Martin}{Martin}{2019b}]%
        {30}
\bibfield{author}{\bibinfo{person}{Fowler Martin}.}
  \bibinfo{year}{2019}\natexlab{b}.
\newblock \bibinfo{title}{Technical Debt}.
\newblock
\newblock
\urldef\tempurl%
\url{https://martinfowler.com/bliki/TechnicalDebt.html}
\showURL{%
\tempurl}


\bibitem[\protect\citeauthoryear{Martini, Bosch, and Chaudron}{Martini
  et~al\mbox{.}}{2015}]%
        {39}
\bibfield{author}{\bibinfo{person}{Antonio Martini}, \bibinfo{person}{Jan
  Bosch}, {and} \bibinfo{person}{Michel Chaudron}.}
  \bibinfo{year}{2015}\natexlab{}.
\newblock \showarticletitle{Investigating Architectural Technical Debt
  Accumulation and Refactoring over Time}.
\newblock \bibinfo{journal}{\emph{Inf. Softw. Technol.}} \bibinfo{volume}{67},
  \bibinfo{number}{C} (\bibinfo{date}{Nov.} \bibinfo{year}{2015}),
  \bibinfo{pages}{237–253}.
\newblock
\showISSN{0950-5849}
\urldef\tempurl%
\url{https://doi.org/10.1016/j.infsof.2015.07.005}
\showURL{%
\tempurl}


\bibitem[\protect\citeauthoryear{Molnar and Motogna}{Molnar and
  Motogna}{2017}]%
        {3}
\bibfield{author}{\bibinfo{person}{Arthur Molnar} {and} \bibinfo{person}{Simona
  Motogna}.} \bibinfo{year}{2017}\natexlab{}.
\newblock \showarticletitle{Discovering Maintainability Changes in Large
  Software Systems}. In \bibinfo{booktitle}{\emph{Proceedings of the 27th
  International Workshop on Software Measurement and 12th International
  Conference on Software Process and Product Measurement}} (Gothenburg, Sweden)
  \emph{(\bibinfo{series}{IWSM Mensura '17})}. \bibinfo{publisher}{ACM},
  \bibinfo{address}{New York, NY, USA}, \bibinfo{pages}{88--93}.
\newblock
\showISBNx{978-1-4503-4853-9}
\urldef\tempurl%
\url{https://doi.org/10.1145/3143434.3143447}
\showDOI{\tempurl}


\bibitem[\protect\citeauthoryear{Molnar and Motogna}{Molnar and
  Motogna}{2020}]%
        {4}
\bibfield{author}{\bibinfo{person}{Arthur{-}Jozsef Molnar} {and}
  \bibinfo{person}{Simona Motogna}.} \bibinfo{year}{2020}\natexlab{}.
\newblock \showarticletitle{Longitudinal Evaluation of Open-Source Software
  Maintainability}. In \bibinfo{booktitle}{\emph{Proceedings of the 15th
  International Conference on Evaluation of Novel Approaches to Software
  Engineering (ENASE)}}. INSTICC, \bibinfo{publisher}{SciTePress},
  \bibinfo{pages}{120--131}.
\newblock
\showISBNx{978-989-758-421-3}


\bibitem[\protect\citeauthoryear{Molnar., Neamţu., and Motogna.}{Molnar.
  et~al\mbox{.}}{2019}]%
        {7}
\bibfield{author}{\bibinfo{person}{Arthur{-}Jozsef Molnar.},
  \bibinfo{person}{Alexandra Neamţu.}, {and} \bibinfo{person}{Simona
  Motogna.}} \bibinfo{year}{2019}\natexlab{}.
\newblock \showarticletitle{Longitudinal Evaluation of Software Quality Metrics
  in Open-Source Applications}. In \bibinfo{booktitle}{\emph{Proceedings of the
  14th International Conference on Evaluation of Novel Approaches to Software
  Engineering - Volume 1: ENASE,}}. INSTICC, \bibinfo{publisher}{SciTePress},
  \bibinfo{pages}{80--91}.
\newblock
\showISBNx{978-989-758-375-9}
\urldef\tempurl%
\url{https://doi.org/10.5220/0007725600800091}
\showDOI{\tempurl}


\bibitem[\protect\citeauthoryear{Molnar}{Molnar}{2020}]%
        {50}
\bibfield{author}{\bibinfo{person}{Arthur-Jozsef Molnar}.}
  \bibinfo{year}{2020}\natexlab{}.
\newblock \showarticletitle{{Data Set for Evaluation of Technical Debt in
  Open-Source Software}}.
\newblock  (\bibinfo{date}{7} \bibinfo{year}{2020}).
\newblock
\urldef\tempurl%
\url{https://doi.org/10.6084/m9.figshare.c.4990778}
\showURL{%
\tempurl}


\bibitem[\protect\citeauthoryear{{Nayebi}, {Cai}, {Kazman}, {Ruhe}, {Feng},
  {Carlson}, and {Chew}}{{Nayebi} et~al\mbox{.}}{2019}]%
        {51}
\bibfield{author}{\bibinfo{person}{M. {Nayebi}}, \bibinfo{person}{Y. {Cai}},
  \bibinfo{person}{R. {Kazman}}, \bibinfo{person}{G. {Ruhe}},
  \bibinfo{person}{Q. {Feng}}, \bibinfo{person}{C. {Carlson}}, {and}
  \bibinfo{person}{F. {Chew}}.} \bibinfo{year}{2019}\natexlab{}.
\newblock \showarticletitle{A Longitudinal Study of Identifying and Paying Down
  Architecture Debt}. In \bibinfo{booktitle}{\emph{2019 IEEE/ACM 41st
  International Conference on Software Engineering: Software Engineering in
  Practice (ICSE-SEIP)}}. \bibinfo{pages}{171--180}.
\newblock


\bibitem[\protect\citeauthoryear{Nugroho, Visser, and Kuipers}{Nugroho
  et~al\mbox{.}}{2011}]%
        {41}
\bibfield{author}{\bibinfo{person}{Ariadi Nugroho}, \bibinfo{person}{Joost
  Visser}, {and} \bibinfo{person}{Tobias Kuipers}.}
  \bibinfo{year}{2011}\natexlab{}.
\newblock \showarticletitle{An empirical model of technical debt and interest}.
  In \bibinfo{booktitle}{\emph{Proceedings of the 2nd Workshop on Managing
  Technical Debt (MTD '11)}}. \bibinfo{pages}{1--8}.
\newblock


\bibitem[\protect\citeauthoryear{Runeson and H{\"o}st}{Runeson and
  H{\"o}st}{2008}]%
        {17}
\bibfield{author}{\bibinfo{person}{Per Runeson} {and} \bibinfo{person}{Martin
  H{\"o}st}.} \bibinfo{year}{2008}\natexlab{}.
\newblock \showarticletitle{Guidelines for conducting and reporting case study
  research in software engineering}.
\newblock \bibinfo{journal}{\emph{Empirical Software Engineering}}
  \bibinfo{volume}{14} (\bibinfo{year}{2008}), \bibinfo{pages}{131--164}.
\newblock


\bibitem[\protect\citeauthoryear{{SIG Group}}{{SIG Group}}{2018}]%
        {48}
\bibfield{author}{\bibinfo{person}{{SIG Group}}.}
  \bibinfo{year}{2018}\natexlab{}.
\newblock \bibinfo{title}{{SIG} Quality Model 2018}.
\newblock
\newblock
\urldef\tempurl%
\url{https://www.softwareimprovementgroup.com/resources/sig-quality-model-2018-now-available/}
\showURL{%
\tempurl}


\bibitem[\protect\citeauthoryear{SonarSource}{SonarSource}{2020}]%
        {1}
\bibfield{author}{\bibinfo{person}{SonarSource}.}
  \bibinfo{year}{2020}\natexlab{}.
\newblock \bibinfo{title}{SonarQube platform user guide}.
\newblock
  \bibinfo{howpublished}{\url{https://docs.sonarqube.org/latest/user-guide/}}.
\newblock


\bibitem[\protect\citeauthoryear{Stre\v{c}ansk\'{y}, Chren, and
  Rossi}{Stre\v{c}ansk\'{y} et~al\mbox{.}}{2020}]%
        {53}
\bibfield{author}{\bibinfo{person}{Peter Stre\v{c}ansk\'{y}},
  \bibinfo{person}{Stanislav Chren}, {and} \bibinfo{person}{Bruno Rossi}.}
  \bibinfo{year}{2020}\natexlab{}.
\newblock \showarticletitle{Comparing Maintainability Index, SIG Method, and
  SQALE for Technical Debt Identification}. In
  \bibinfo{booktitle}{\emph{Proceedings of the 35th Annual ACM Symposium on
  Applied Computing}} (Brno, Czech Republic) \emph{(\bibinfo{series}{SAC
  ’20})}. \bibinfo{publisher}{Association for Computing Machinery},
  \bibinfo{address}{New York, NY, USA}, \bibinfo{pages}{121–124}.
\newblock
\showISBNx{9781450368667}
\urldef\tempurl%
\url{https://doi.org/10.1145/3341105.3374079}
\showURL{%
\tempurl}


\bibitem[\protect\citeauthoryear{Walkinshaw and Minku}{Walkinshaw and
  Minku}{2018}]%
        {38}
\bibfield{author}{\bibinfo{person}{Neil Walkinshaw} {and}
  \bibinfo{person}{Leandro Minku}.} \bibinfo{year}{2018}\natexlab{}.
\newblock \showarticletitle{Are 20\% of Files Responsible for 80\% of
  Defects?}. In \bibinfo{booktitle}{\emph{Proceedings of the 12th ACM/IEEE
  International Symposium on Empirical Software Engineering and Measurement}}
  (Oulu, Finland) \emph{(\bibinfo{series}{ESEM ’18})}.
  \bibinfo{publisher}{Association for Computing Machinery},
  \bibinfo{address}{New York, NY, USA}, \bibinfo{pages}{1--10}.
\newblock
\showISBNx{9781450358231}
\urldef\tempurl%
\url{https://doi.org/10.1145/3239235.3239244}
\showURL{%
\tempurl}


\bibitem[\protect\citeauthoryear{Ward}{Ward}{1992}]%
        {29}
\bibfield{author}{\bibinfo{person}{Cunningham Ward}.}
  \bibinfo{year}{1992}\natexlab{}.
\newblock \showarticletitle{The WyCash Portfolio Management System}.
\newblock \bibinfo{journal}{\emph{SIGPLAN OOPS Mess.}} \bibinfo{volume}{4},
  \bibinfo{number}{2} (\bibinfo{year}{1992}), \bibinfo{pages}{29--30}.
\newblock


\bibitem[\protect\citeauthoryear{Welker}{Welker}{2001}]%
        {57}
\bibfield{author}{\bibinfo{person}{Kurt Welker}.}
  \bibinfo{year}{2001}\natexlab{}.
\newblock \showarticletitle{Software Maintainability Index Revisited}.
\newblock \bibinfo{journal}{\emph{Journal of Defense Software Engineering}}
  (\bibinfo{date}{08} \bibinfo{year}{2001}).
\newblock


\bibitem[\protect\citeauthoryear{Yuan and Memon}{Yuan and Memon}{2010}]%
        {6}
\bibfield{author}{\bibinfo{person}{Xun Yuan} {and} \bibinfo{person}{Atif~M.
  Memon}.} \bibinfo{year}{2010}\natexlab{}.
\newblock \showarticletitle{Generating Event Sequence-Based Test Cases Using
  {GUI} Run-Time State Feedback}.
\newblock \bibinfo{journal}{\emph{IEEE Transactions on Software Engineering}}
  \bibinfo{volume}{36}, \bibinfo{number}{1} (\bibinfo{year}{2010}),
  \bibinfo{pages}{81--95}.
\newblock
\showISSN{0098-5589}
\urldef\tempurl%
\url{http://doi.ieeecomputersociety.org/10.1109/TSE.2009.68}
\showURL{%
\tempurl}


\end{thebibliography}

\end{document}